\documentclass[sigconf,natbib=true,anonymous=false]{acmart}

\usepackage{newfloat}
\usepackage{listings}
\usepackage{algorithm}
\usepackage{verbatim}
\usepackage{adjustbox}
\usepackage{multirow}
\usepackage{booktabs}
\newtheorem{definition}{Definition}
\newtheorem{theorem}{Theorem}

\usepackage{algorithmic}
\usepackage{graphicx}
\usepackage{textcomp}
\usepackage{xcolor}
\usepackage{comment}
\usepackage[table]{xcolor}

\AtBeginDocument{%
  }

\setcopyright{acmlicensed}
\copyrightyear{2018}
\acmYear{2018}
\acmDOI{XXXXXXX.XXXXXXX}
\acmConference[Conference acronym 'XX]{Make sure to enter the correct
  conference title from your rights confirmation email}{June 03--05,
  2018}{Woodstock, NY}
\acmISBN{978-1-4503-XXXX-X/18/06}

\begin{document}


\title{A Reinforcement Learning Method to Factual and Counterfactual Explanations for Session-based Recommendation}


  \author{Han Zhou}
\email{zhouhan@163.sufe.edu.cn}
\affiliation{%
  \institution{Shanghai University of Finance and Economics}
  \city{Shanghai}
  \country{China}
}

\author{Hui Fang}
\affiliation{%
  \institution{Shanghai University of Finance and Economics}
  \city{Shanghai}
  \country{China}}
\email{fang.hui@mail.shufe.edu.cn}

\author{Zhu Sun}
\affiliation{%
  \institution{Singapore University of Technology and Design}
  \city{Singapore}
  \country{Singapore}
}
\email{zhu\_sun@sutd.edu.sg}

\author{WenTao Hu}
\affiliation{%
  \institution{Shanghai University of Finance and Economics}
  \city{Shanghai}
  \country{China}}
\email{stevenhwt@gmail.com}

\renewcommand{\shortauthors}{Trovato et al.}

\begin{abstract}
Session-based Recommendation (SR) systems have recently achieved considerable success, yet their complex, "black box" nature often obscures why certain recommendations are made. Existing explanation methods struggle to pinpoint truly influential factors, as they frequently depend on static user profiles or fail to grasp the intricate dynamics within user sessions. In response, we introduce FCESR (Factual and Counterfactual Explanations for Session-based Recommendation), a novel framework designed to illuminate SR model predictions by emphasizing both the sufficiency (factual) and necessity (counterfactual) of recommended items. By recasting explanation generation as a combinatorial optimization challenge and leveraging reinforcement learning, our method uncovers the minimal yet critical sequence of items influencing recommendations.
Moreover, recognizing the intrinsic value of robust explanations, we innovatively utilize these factual and counterfactual insights within a contrastive learning paradigm, employing them as high-quality positive and negative samples to fine-tune and significantly enhance SR accuracy. Extensive qualitative and quantitative evaluations across diverse datasets and multiple SR architectures confirm that our framework not only boosts recommendation accuracy but also markedly elevates the quality and interpretability of explanations, thereby paving the way for more transparent and trustworthy recommendation systems.
\end{abstract}



\maketitle

\section{Introduction}

Session-based recommendation (SR) has become integral in industries such as e-commerce and multimedia streaming, where understanding user intent within a single interaction session is crucial \cite{wang2021survey}.
Most existing research on SR aims to enhance recommendation accuracy by employing complex methods like Markov Chains \cite{Stamen2022} and deep learning methods like recurrent neural networks (RNN) \cite{hidasi2015session,quad17HRNN}, attention mechanisms \cite{Liu2018STAMP}, and graph neural networks (GNN) \cite{wu2019session}.  However, these deep learning approaches often lack transparency, making it difficult for users to understand the reasoning behind the recommendations. This opacity can ultimately reduce user experience and satisfaction. Therefore, it is essential to develop SR models that not only prioritize accuracy but also offer meaningful explanations for their recommendations.

Typically, effective explanations should fulfill two key purposes: building trust by enabling a normative evaluation of algorithmic decisions, and providing actionable recourse for users to improve future outcomes \cite{selbst2018intuitive,2021generalexp}. To achieve these objectives, existing methods in explainable artificial intelligence (XAI) can be broadly categorized along two dimensions: the timing of explanation generation, spanning post-hoc explainable methods and model-intrinsic methods \cite{mittelstadt2019explaining}, and the dependency on model internals, ranging from model-dependent to model-agnostic approaches \cite{2018expsurvey}.

\begin{figure}[htbp] 
\centering
\includegraphics[width=0.5\textwidth]
{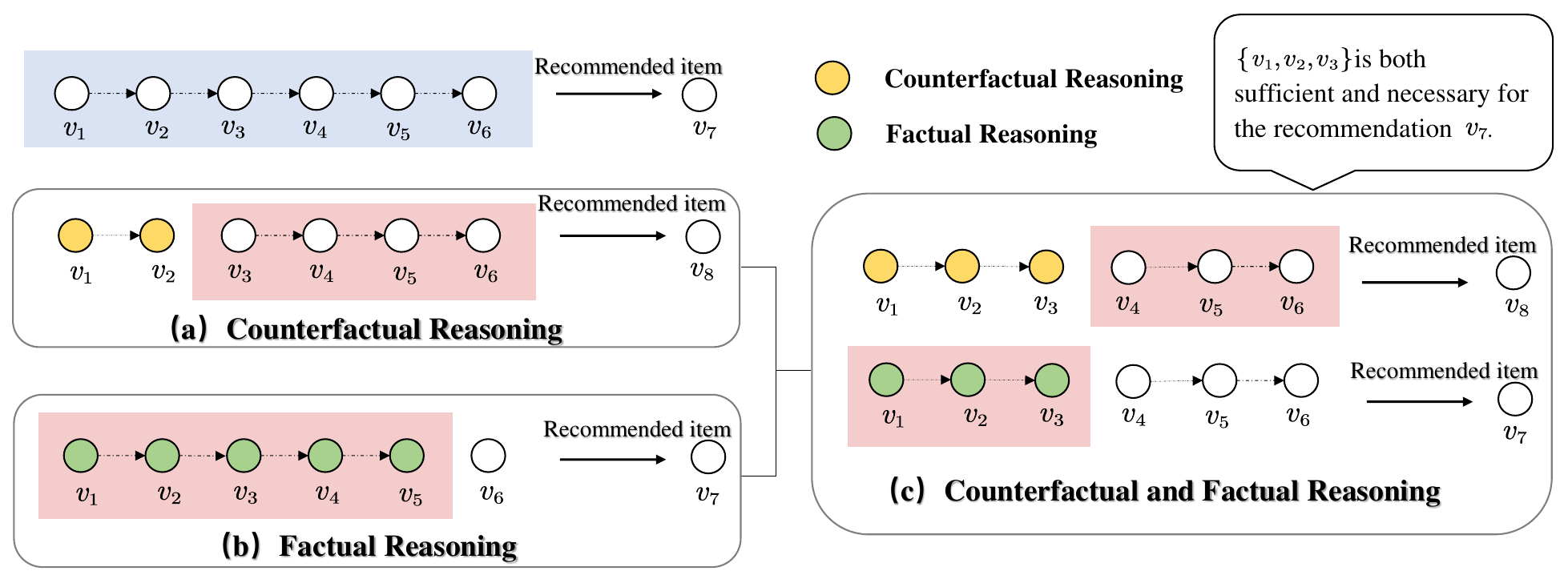} 
\captionsetup{font={small}}
\vspace{-0.15in}
\caption{Definitions of factual and counterfactual explanations. Circles represent items in a session, where yellow indicates key counterfactual items and green indicates key factual items:
(a) counterfactual reasoning: removing $v_1$ and $v_2$ changes the recommendation from $v_7$ to $v_8$; (b) factual reasoning: the presence of $v_1$ to $v_5$ is sufficient to recommend $v_7$; and (c) counterfactual and factual reasoning: \{$v_1$, $v_2$, $v_3$\} are both sufficient and necessary conditions for recommending $v_7$.} 
\label{fig:definition} 
\vspace{-0.15in}
\end{figure}

In sequential data scenarios, recent efforts have focused on improving explainability through various approaches. These include the integrated path and embedding method \cite{2019houESRKG} and the knowledge graph-based approach RSL-GRU \cite{Cui2022RSL}, both of which are post-hoc methods, as well as the model-intrinsic path-based framework REKS \cite{wu2023generic}. However, these existing methods face two significant limitations: they often require additional data resources such as knowledge graphs and fail to provide true explanations. 
A \emph{true explanation}, as defined by \cite{xu2021learning,peake2018kdd}, identifies the specific components of past inputs that significantly influence the recommendation outcome, thereby clarifying which items drive the system's decision. By capturing the causal relationships between inputs and outputs, true explanations help avoid generating incorrect or misleading interpretations.
Therefore, in this work, we focus on generating true explanations for SR through a post-hoc, model-agnostic framework.



\emph{Counterfactual-thinking based explanations} have been widely applied in recommender systems (RSs) \cite{tan2021counterfactual, gao2024causal}. These approaches aim to provide true explanations by identifying minimal perturbations that can alter the recommended outputs. However, when applied to sequential data, current methods are subject to two notable challenges. First, most existing explanations primarily target non-sequential data. Second, while some progress has been made for sequential data—such as \cite{xu2021learning}, which explored true explanations for sequential recommendations by identifying influential items through causal dependencies using counterfactual data, there are still several weaknesses: (1) It lacks practical insights that could be translated into actionable interventions (i.e., modifying user interactions or preferences to influence recommendation outcomes). (2) The variational autoencoder (VAE) they employed for perturbation shows limitations in capturing sequential patterns effectively. (3) Their approximate methods for item substitution in recommendation lists may lead to potential biases.


While counterfactual reasoning identifies only the most critical information where removing key sub-items would alter recommendation predictions, \emph{factual reasoning} retains sufficient sub-items to maintain the same prediction, though potentially including redundancies. Based on the definitions of necessity and sufficiency conditions ($X$ is necessary for $Y$ if and only if $Y \rightarrow X$ and $X$ is sufficient for $Y$ if and only if $X \rightarrow Y$), counterfactual reasoning corresponds to necessary conditions, as $Y \rightarrow X$ is equivalent to $ \neg X \rightarrow \neg Y$ (removing sub-items changes the result). In contrast, factual reasoning represents sufficient conditions, as retaining sub-items preserves the result. Combining both approaches can balance necessity and sufficiency, as purely sufficient explanations may include redundancies while purely necessary explanations might miss important items. This balance has been frequently explored in explaining machine learning models \cite{2021generalexp} and GNN-based recommendation models \cite{tan2022learning, chen2022grease}, yet it remains unexplored in the context of SR scenarios.

In this context, we first formally define factual and counterfactual reasoning in SR. As illustrated in Figure \ref{fig:definition}, intuitively, the \emph{sufficient} (i.e., factual) conditions for an item to be recommended to a user are the set of items that enable that item to be included in the final list of top-$K$ recommended items. A factual explanation might be: ``You were recommended item $Y$ because you interacted with item $X$". In contrast, the \emph{necessary} (i.e., counterfactual) conditions are the set of items without which that item would not appear in the final list of top-$K$ recommended items. A counterfactual explanation example could be: ``If you had not interacted with item $X$, you would not have been recommended item $Y$".
Generating sufficient and necessary model-agnostic explanations for existing SR models is valuable but also faces two major challenges: (1) Effectively modeling sequential data to capture dependencies between items, rather than treating them as isolated entities. (2) Balancing factual and counterfactual objectives (i.e., sufficiency and necessity) in both modeling and evaluation ensures that explanations account for both perspectives effectively.

On the other hand, the derived factual and counterfactual explanations may potentially enhance recommendation accuracy by providing high-quality positive and negative signals of sequential modeling. These derived explanations can be viewed as a form of data augmentation, effectively enriching the training data. How to effectively leverage these explanations, beyond merely increasing system transparency, remains an open research question. 
Exploring this problem could lead to better utilization of explanatory information, ultimately helping to maximize the use of high-quality samples for simultaneously improving both recommendation accuracy and explanation quality.

Thus, we propose FCESR, a model-agnostic framework based on reinforcement learning (RL) that integrates both factual and counterfactual reasoning to enhance the explainability of SR models. Furthermore, we leverage contrastive learning to utilize the generated explanations for improving recommendation effectiveness. Our main contributions are as follows.
(1) We formulate the problem of generating factual and counterfactual explanations as a combinatorial optimization problem and provide a systematic analysis of its solution approach.
(2) We effectively capture sequential information using RL to thoroughly explore item dependencies within sessions. Additionally, we design a reward mechanism that carefully balances necessity and sufficiency by considering both counterfactual and factual explanation objectives.
(3) We apply the contrastive learning technique to fine-tune the corresponding SR model with the generated explanations, which not only provides high-quality explanations but also improves recommendation effectiveness.
(4) We validate the effectiveness of our framework in improving recommendation accuracy and explanation quality through extensive experiments on three datasets, utilizing five state-of-the-art non-explainable SR models.

\section{Related Work}

Our work is related to three areas of research: explainable recommendation, factual and counterfactual reasoning, and contrastive learning.

\subsection{Explainable Recommendation}
Explainable Recommendation aims to enhance user satisfaction and system transparency~\cite{zhang2020explainable}, typically classified into model-dependent and model-agnostic methods. 
The former \cite{shulman2020meta,lin2000collaborative,vig2009} integrates explanations directly into the recommendation process but often increases complexity and may reduce overall explainability. In contrast, the latter \cite{verma2020counterfactual,peake2018kdd} generates explanations independently, offering greater flexibility.
Some studies have explored explainability for sequential data \cite{2019houESRKG,Cui2022RSL,xu2021learning}, but mainly focused on sequential recommendation with explicit user information, rather than SR where users are anonymous. Additionally, many of them require supplementary data, such as knowledge graph. In SR, REKS~\cite{wu2023generic} introduces a path-based generic explainable framework, though it still requires extra knowledge graph information. To summarize, while few studies have explored explainable SR, most of them for sequential data depend on side information, and fail to provide true explanations (identifying causally influential inputs) and actionable explanations (enabling users to adjust interactions to change outcomes), limiting their effectiveness in enhancing transparency.

\subsection{Factual and Counterfactual Explanation}
Factual explanations, like LIME \cite{ribeiro2016should} and SHAP \cite{lundberg2017unified}, highlight key input components that sustain a model's prediction by removing less important information. In recommendation, especially those utilizing GNNs \cite{chen2022grease}, they enhance interpretability by identifying crucial interactions that influence recommendations.
Conversely, counterfactual reasoning focuses on how changes in input can alter model predictions. It has been applied to  fairness \cite{ge2022explainable, kusner2017counterfactual}, augment training samples \cite{zhang2021causerec, chen2022data}, causality learning \cite{song2023counterfactual} and  casual explanation \cite{xu2021learning}. 
%
While factual and counterfactual reasoning are widely applied, there is a need to combine their strengths to identify sufficient and necessary conditions to explain SR with sequential properties, and to use these explanations to improve recommendation accuracy.

\subsection{Contrastive Learning}
Recent advancements in recommendation have utilized contrastive learning (CL) to enhance model performance and robustness.
In non-sequential recommendations,
CL leverages static features, improving data representations without relying on temporal dynamics \cite{zhou2023contrastive,Yu2022SelfSupervisedLF}. 
In sequential recommendations, studies such as \cite{zhang2024tois,Yang2023TheWebConf,Xie2022ContrastiveLF,Chen2022IntentCL, Liu2021ContrastiveSS} demonstrate the application of CL to sequential data, where multiple temporal views of user interactions are contrasted to refine the prediction models. In our study, we, for the first time, exploit factual and counterfactual explanations to provide high-quality positive and negative samples, which enhance the accuracy of SR models by taking the advantages of contrastive learning.

\section{Preliminary}


In the context of SR, let $\mathcal{V} = \{v_1, v_2, \ldots, v_{|\mathcal{V}|}\}$ denote the item set. An arbitrary session $\mathcal{S}$ is expressed as a sequence of items $\mathcal{S} = \{v_{1}^{\mathcal{S}}, v_{2}^{\mathcal{S}}, \ldots, v_{|{\mathcal{S}}|}^{\mathcal{S}}\}$ with length $|{\mathcal{S}}|$. Here, $v_{j}^{\mathcal{S}} \in \mathcal{V}$ is the $j$-th interacted item, such as clicked, watched, or purchased item. 
Suppose that for an anonymous session ${\mathcal{S}}$, the recommended item is $i^*$. 
We define the explanations as follows:

\begin{definition}
\textbf{Factual Explanation (FE).}
A FE for the recommended item $i^*$ is represented by an item set for factual explanation $\mathcal{S}^+$. $\mathcal{S}^+$ captures the sufficient condition for recommending item $i^*$, highlighting the key user interactions that directly lead to this recommendation.
Formally, when $\mathcal{S}^{+}$ is fed into the recommendation model, $i^*$ consistently appears in the top-$K$ recommendation list.
\end{definition}

\begin{definition}
\noindent\textbf{Counterfactual Explanation (CFE).}
A CFE for the recommended item $i^*$ is represented by an item set for counterfactual explanation $\mathcal{S}^-$. $\mathcal{S}^-$ indicates the necessary condition for recommending item $i^*$.
Formally, when $\mathcal{S} \setminus \mathcal{S}^{-}$ ($\mathcal{S}$ excluding $\mathcal{S}^-$) is fed into the recommendation model, $i^*$ is notably absent from the top-$K$ recommendation list.
\end{definition}

\begin{definition}
\noindent\textbf{Factual and Counterfactual Explanation (CF$\boldsymbol{^2}$)}.
A CF$\boldsymbol{^2}$ encapsulates both sufficient and necessary conditions for recommending item $i^{*}$, represented by $\mathcal{S}^*$, a subset of the session $\mathcal{S}$. $\mathcal{S}^*$ serves as the explanation for recommending item $i^*$ and satisfies two conditions: (1) \emph{Sufficiency}: feeding $\mathcal{S}^*$ into the recommendation model results in $i^*$ appearing in the top-$K$ recommendation list; (2) \emph{Necessity}: inputting $ \mathcal{S} \setminus \mathcal{S}^*$ prevents $i^*$ from appearing in the top-$K$ list. This dual nature of $\mathcal{S}^{*}$ means that it contains the crucial interactions leading to the recommendation of $i^{*}$, and removing any items in $\mathcal{S}^{*}$ would stop $i^{*}$ from being recommended. 
\end{definition}

\subsection{Task Formulation}
We need to tackle two main tasks. The first is to provide explanations for recommendation results of non-explainable SR models by leveraging both factual and counterfactual conditions. The second is to enhance recommendation accuracy of these SR models by exploiting the derived explanations.

For each session $\mathcal{S}$, we use a SR model $f_{R}$ (e.g., NARM \cite{li2017neural}) to generate embeddings for the entire session and individual items within it. $f_{R}$ produces a top-$K$ recommendation list for the next-item prediction task:
\begin{equation}
\footnotesize
    \mathbf{e}_\mathcal{S}, \mathbb{I}_\mathcal{S}^K =f_{R} (\mathcal{S}),
\end{equation}
where $\mathbf{e}_\mathcal{S}$ and $\mathbb{I}_\mathcal{S}^K$ represent the session embedding and the top-$K$ recommendation list for session $\mathcal{S}$, respectively.

Given session $\mathcal{S}$, 
we aim to design an \emph{explanation model} that generates both factual and counterfactual explanations (i.e. sub-sessions) for the recommended item $i^*$ by the SR model that either support or challenge the recommendation results.
Formally, this explanation model is defined as a combinatorial optimization problem, referred to as the \textit{Min-Explanation} problem:
\begin{equation}\label{eq:optimization}
\footnotesize
\begin{array}{ll}
\operatorname{minimize} & C(M) \\
\text{s.t.,} & i^* \in \mathbb{I}_{\mathcal{S} \odot M}^K \\
              & i^* \notin \mathbb{I}_{\mathcal{S}  \backslash (\mathcal{S} \odot M) }^K \\
              & M_{i} \in\{0,1\}, \quad \forall i=1,2, \ldots,|\mathcal{S}|.
\end{array}
\end{equation}
Here, $M$ is a binary vector $ M \in \{0,1\}^{|\mathcal{S}|}$, where each element $M_i$ indicates whether the corresponding item $v_i^\mathcal{S}$ in $\mathcal{S}$ is included in the explanation ($M_i = 1$) or not ($M_i = 0$). $\mathcal{S} \odot M$ denotes the element-wise product of $\mathcal{S}$ and $M$ ($\mathcal{S}$ is treated as a vector consisting of items in the session), representing the factual and counterfactual explanation $\mathcal{S}^*$ for session $\mathcal{S}$, while $\mathcal{S} \backslash (\mathcal{S} \odot M)$ represents the session with explained items removed. $C(M)$ is explanation complexity, defined as follows:
\begin{definition}
\textbf{Explanation Complexity.} We define the complexity $C(M)$ as the $L_0$ norm of $M$, i.e., $C(M) = \|M\|_0=\sum_{i=1}^{|\mathcal{S}|} M_{i}$, which represents the number of non-zero elements in $M$. Minimizing this complexity promotes more concise and interpretable explanations.
\end{definition}

The optimization problem (see Equation \ref{eq:optimization}) incorporates both factual and counterfactual conditions: \textbf{(1) Factual Condition}. 
 The recommended item  ${i}^{*}$  should appear in the top-$K$ recommendations when considering only the items selected by $M$: ${i}^{*} \in \mathbb{I}_{\mathcal{S} \odot M}^{K}$. This ensures that the items selected in $\mathcal{S} \odot M$ are sufficient for  ${i}^{*}$ to be recommended. \textbf{(2) Counterfactual Condition}. The recommended item ${i}^{*}$ should not appear in the top-$K$  recommendations when the selected items $\mathcal{S} \odot M$ are removed from the session: ${i}^{*} \notin \mathbb{I}_{\mathcal{S} \backslash (\mathcal{S} \odot M)}^{K}$. This indicates that the absence of the selected items would prevent item ${i}^{*}$ from being recommended.

\begin{figure*}[t] 
\centering
\includegraphics[width=0.9\textwidth]{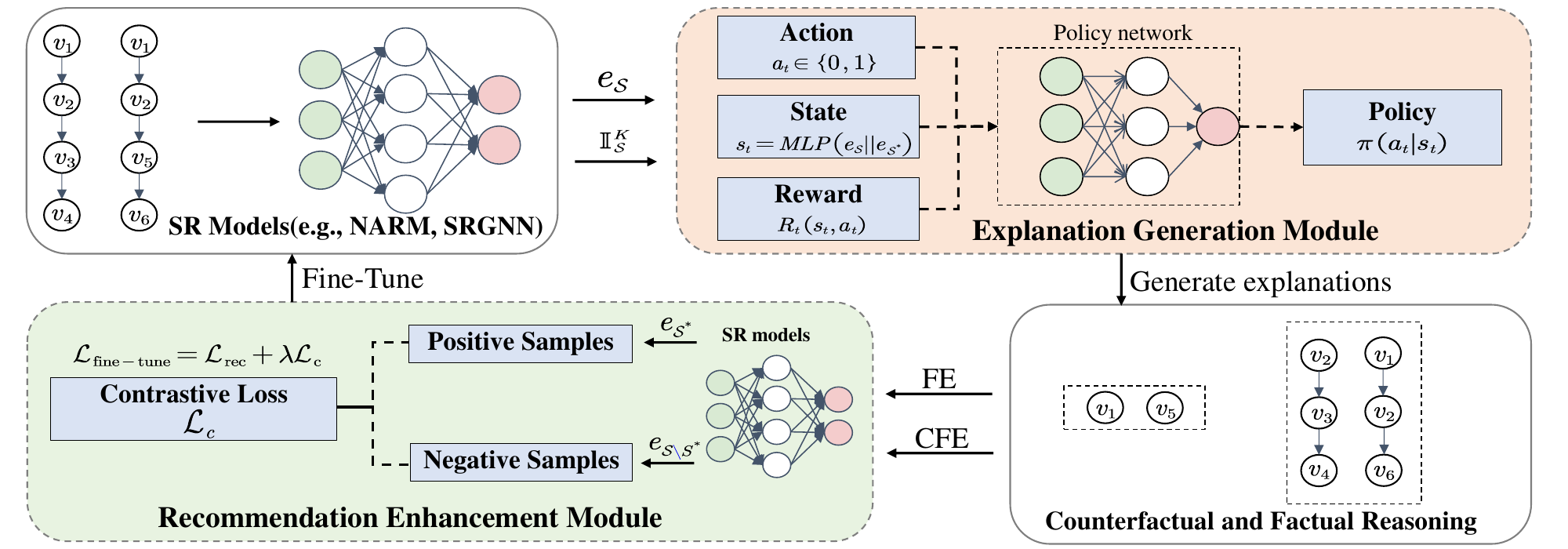} 
\vspace{-0.1in}
\caption{Overview of the FCESR framework.} 
\label{fig:overview} 
\vspace{-0.15in}
\end{figure*}

\subsection{Analysis of the Min-Explanation Problem}
We analyze the computational complexity of the \textit{Min-Explanation} optimization problem and justify using reinforcement learning (RL) as an effective solution.

\smallskip \noindent \textbf{Time Complexity Analysis.} For a session  $\mathcal{S}$ with  $|\mathcal{S}|$ items, the solution space comprises all possible subsets of these items, resulting in $2^{|\mathcal{S}|}$ combinations. The number of potential solutions grows exponentially with $|\mathcal{S}|$, making any algorithm that exhaustively searches for the optimal solution incur exponential time complexity. This renders the problem computationally infeasible for large $|\mathcal{S}|$. We demonstrate the computational hardness of the \textit{Min-Explanation} problem by reducing it from the NP-Complete \textit{MIN-FEATURES} problem \cite{davies1994np}.
\begin{theorem}
    The   \textit{Min-Explanation} problem is NP-Hard.
\end{theorem}
\vspace{-0.1in}
\begin{proof}
   To prove that this combinatorial optimization problem is NP-hard, we reduce the known NP-Complete problem \textit{MIN-FEATURES} to our problem. Specifically, we map each feature in the \textit{MIN-FEATURES} problem to an item $v_{i}^\mathcal{S}$ in our session $\mathcal{S}$, and feature selection corresponds to selecting elements with the binary vector $M$. The goal of \textit{MIN-FEATURES} is to find a minimal subset of features of size $n$. This aligns with our optimization problem, where we aim to minimize the explanation complexity $C(M)$ while satisfying the factual and counterfactual conditions. Through this mapping, a solution to the \textit{MIN-FEATURES} problem can be transformed into a solution to our problem, and vice versa, thereby proving their equivalence. Since \textit{MIN-FEATURES} is NP-Complete, this reduction demonstrates that our combinatorial optimization problem is also NP-hard. 
\end{proof}

\noindent \textbf{Using Reinforcement Learning to Solve the Min-Explanation Problem.} In recent years, novel approaches leveraging deep reinforcement learning (RL) have emerged as effective tools for addressing combinatorial optimization problems \cite{mazyavkina2021reinforcement}. These methods have shown remarkable success in solving classical problems such as the Traveling Salesman Problem (TSP) \cite{NCO2017}.
Building upon these advances, we propose a RL-based solution for the \textit{Min-Explanation} problem within a Markov Decision Process (MDP) framework. Our choice is motivated by three key advantages.(1) Selecting the optimal subset of items for explanation is inherently a combinatorial optimization problem. Traditional methods struggle with the exponential search space, especially for longer sessions, making traditional exhaustive or heuristic methods computationally infeasible. RL, on the other hand, can effectively navigate this space by learning a policy for sequential decision-making.
(2) SR models deal with user interaction sequences. RL is particularly well-suited to capture the temporal dynamics and dependencies within a session.
(3) The \textit{Min-Explanation} problem involves balancing multiple objectives: minimizing explanation complexity while satisfying both factual and counterfactual conditions. RL enables the design of a reward function that effectively balances these competing objectives. 
Beyond generating explanations, we also aim to use these explanations to further enhance the accuracy of SR model  $f_{R}$.

\section{Our FCESR Framework}

We present our novel FCESR framework, designed to generate factual and counterfactual explanations for any SR model. Figure \ref{fig:overview} illustrates its architecture, which consists of two main components: \emph{Explanation Generation Module} and \emph{Recommendation Enhancement Module}. Specifically, the former module
provides both factual and counterfactual explanations for recommended items given a SR model.
It models the task within a MDP framework, utilizing RL to efficiently capture sequential information and address the combinatorial optimization challenge.
The latter module further leverages the generated explanations to improve the effectiveness of the SR model.
%
%
The overall framework is model-agnostic and can be applied to any SR model. This flexibility ensures that our approach can be widely adopted across different SR systems, enhancing both their interpretability and accuracy.

\subsection{Explanation Generation Module}
This module uses MDP to generate CF$\boldsymbol{^2}$ for SR, providing sufficient and necessary conditions. We define the MDP environment as a quadruple \( (S, A, T, R) \): (1) \( S \) is the state space, where each state \( s_t \) represents the current agent's information; (2) \( A \) is the action space, involving the selection of items to include in the explanations; (3) \( {T}: S \times A \rightarrow S \) is the state transition function, describing how the state changes based on the action taken; and (4) \( R: {S} \times {A} \rightarrow \mathbb{R} \) is the reward function, evaluating the effectiveness of actions by how well the added explanations align with user expectations and enhance their understanding of the recommendations. The policy $\pi(a_t | s_t)$ guides action selection and aims to maximize cumulative reward by choosing the optimal items for explanation. The explanation policy is continuously refined through a process of learning and adaptation, leveraging feedback to improve the relevance and impact of the generated explanations.

\noindent\textbf{State.}
We initialize the CF$\boldsymbol{^2}$ $\mathcal{S}^{*}$ as $\emptyset$. For the sufficient condition (factual condition), we compute the embeddings and top-$K$ recommendation list as $\mathbf{e}_{\mathcal{S}^*},\mathbb{I}_{\mathcal{S}^*}^K=f_{R} (\mathcal{S}^*)$, where $\mathbf{e}_{\mathcal{S}^*}=\mathbf{0}^d$ if $\mathcal{S}^*$ is empty. For the necessary condition (counterfactual condition), we calculate $\mathbf{e}_{\mathcal{S} \setminus \mathcal{S}^*},\mathbb{I}_{\mathcal{S} \setminus \mathcal{S}^{*}}^K=f_{R} (\mathcal{S} \setminus \mathcal{S}^*)$. 
At step $t$,  the state $ s_t = \text{MLP} (\mathbf{e}_\mathcal{S} \| \mathbf{e}_{\mathcal{S}^*}  )$ is represented by the output of a multi-layer perceptron (MLP), which integrates the embeddings of the current session $\mathbf{e}_\mathcal{S}$ and the explanation $ \mathbf{e}_{\mathcal{S}^*}$.

\noindent\textbf{Action.}
The action space involves deciding whether to include an item in $\mathcal{S}^*$. At each step $t$, the action $a_t \in \{0, 1\}$ is determined by a decision-making model.
During the process, each item in session $\mathcal{S}$ is evaluated at each step $t$: if $a_t = 1$, the current item $v_t^{\mathcal{S}}$ is selected and added to the explanation $\mathcal{S}^*$, updating it to $\mathcal{S}^*_{[t]} = [\mathcal{S}^*_{[1:t-1]}, v_t^{\mathcal{S}}]$; if $a_t = 0$,  the current item $v_t^{\mathcal{S}}$ is not selected for the explanation. The action selection process continues until all items in the session have been traversed ($t = |\mathcal{S}|$).

\noindent\textbf{Agent.}
The policy network, modeled as a MLP, integrates both session and explanation information. On this basis, a sigmoid function is applied to compute the probability of selecting the next item, following the approach suggested by recent literature \cite{wu2023generic}.
The policy \(\pi(a_t \mid s_t)\) guiding action selection is defined as:
\begin{equation*}\footnotesize
    \pi(a_t \mid s_t) = \sigma(A_t \odot \mathbf{W} s_t),
\end{equation*}
where \(\sigma(\cdot)\) is the sigmoid function,\( \mathbf{W} \) is a trainable parameter matrix,  and \( A_t \) is a vector representing the probability of each action. As this is a binary classification problem, the action \( a_t \) is determined based on the output probability:
\begin{equation*}\footnotesize
    a_t = \begin{cases} 
1 & \text{if } \pi(a_t=1 \mid s_t) > 0.5 \\
0 & \text{otherwise .}
\end{cases}
\end{equation*}
In this way, the policy network effectively decides the action at each step $t$, thereby facilitating the generation of high-qualityCF$\boldsymbol{^2}$.

\noindent\textbf{Reward.}
The reward mechanism $R$ is designed to generate minimal sufficient and necessary conditions for recommendations. Consider the scenario of explaining next-item prediction, and $i^*\in \mathbb{I}_{\mathcal{S}}^K$ represents the next item recommended based on the current session $\mathcal{S}$. The reward function comprises four components, each assessing a different aspect of the explanation's effectiveness:

\noindent(1) \textbf{Factual Reward ($ R_{fe}$)}: it evaluates the sufficiency of the explanation by rewarding the model when the recommended item $i^*$ is included in the top-$K$ list generated from $\mathcal{S}^*$.
    \begin{equation}
    \footnotesize
    R_{fe}(s_t, a_t) = \begin{cases} 
    1 & \text{if } i^* \in \mathbb{I}_{\mathcal{S}^*}^K \\
    0 & \text{if } i^* \notin \mathbb{I}_{\mathcal{S}^*}^K ,
    \end{cases}
    \end{equation}
(2) \textbf{Counterfactual Reward ($R_{cfe}$)}: it evaluates the necessity of the explanation by assessing the impact of excluded items. It rewards the model when $i^*$ is not in the top-$K$ list generated from $\mathcal{S} \setminus \mathcal{S}^*$:
\begin{equation}\footnotesize
    R_{cfe}(s_t, a_t) = \begin{cases} 
    0 & \text{if } i^* \in \mathbb{I}_{\mathcal{S} \setminus \mathcal{S}^{*}}^K \\
    1 & \text{if } i^* \notin \mathbb{I}_{\mathcal{S} \setminus \mathcal{S}^{*}}^K ,
    \end{cases}
    \end{equation}
(3) \textbf{Sparsity Reward} (\( R_{sp} \)): it aims to minimize the size of \( \mathcal{S}^* \) to enhance explanation simplicity and focus:
    \begin{equation} \footnotesize
    R_{sp} = \frac{1}{\log (|\mathcal{S}^*| + 2)} , 
    \end{equation}
(4) \textbf{Rank Reward} (\( R_{rank} \)): it is based on the item's rank in the top-$K$ list, promoting higher-ranked items applied to $i^*$ in the explanations:
\begin{equation}\footnotesize
        R_{rank} = \frac{1}{\log (rk{(i^*)} + 2)} .
    \end{equation} 
Here, $rk{(i^*)}$ denotes the rank of item $i^*$ in $\mathbb{I}_{\mathcal{S}^*}^K$. To align with the factual condition, even when $i^{*}$ is no longer included in the top-$K$ recommendation list, we still aim for its ranking to be as high as possible. A lower value of $rk{(i^*)}$ corresponds to a higher reward.

By directly adding them together to balance these rewards, our model is trained to produce explanations that are more concise and meet the sufficient and necessary conditions:
\begin{equation}\footnotesize
       R\left(s_{t}, a_{t}\right)= R_{fe} + R_{cfe}+R_{sp } +R_{rank}.
\end{equation}

\begin{table}[h]
\centering
\setlength{\tabcolsep}{0.3cm} 
\caption{Summarization of the toy example for MDP.}
\vspace{-0.1in}
\begin{tabular}{cccc}
\toprule
Step & Residue Session & $\mathcal{S}^*$   &  Action \\
\midrule
1 & $\_ \rightarrow v_2 \rightarrow v_3 \rightarrow v_4  \rightarrow v_5$ & $ \setminus $ & $1$  \\
\midrule
2 & $v_1 \rightarrow \_ \rightarrow v_3 \rightarrow v_4 \rightarrow v_5$ & $v_1$  & $0$ \\
\midrule
3 & $v_1 \rightarrow  v_2  \rightarrow \_ \rightarrow v_4 \rightarrow v_5 $ & $v_1$  & $0$ \\
\midrule
4 &  $v_1 \rightarrow v_2 \rightarrow v_3  \rightarrow \_ \rightarrow v_5$ & $v_1$  & $1$ \\
\midrule
5 &  $v_1 \rightarrow v_2 \rightarrow v_3  \rightarrow v_4 \rightarrow \_$ & $v_1 \rightarrow v_4$ & $1$ \\
\midrule
Final & - & $v_1 \rightarrow v_4\rightarrow v_5 $  & - \\
\bottomrule
\end{tabular}
\end{table}

\noindent\textbf{A Toy Example of the MDP process.} To illustrate the execution process of our FCESR framework, consider a target session ``$v_1 \rightarrow v_2 \rightarrow v_3 \rightarrow v_4 \rightarrow v_5$''. The task involves traversing the session sequentially. Starting with an empty  CF$\boldsymbol{^2}$ set $\mathcal{S}^*$, the agent decides at each step whether to include the current item in $\mathcal{S}^*$ based on the action $a_t$. 
At \textbf{step 1}, the agent begins with $v_1$ and chooses to include it in the explanation since $a_1 = 1$, updating $\mathcal{S}^*$ to $\{v_1\}$.  
At \textbf{step 2}, the agent moves to $v_2$ but decides not to include it since $a_2 = 0$, leaving $\mathcal{S}^*$ unchanged.  
In \textbf{step 3}, the agent processes $v_3$ and again decides not to include it in the explanation since $a_3 = 0$.  
At \textbf{step 4}, the agent evaluates $v_4$ and chooses to add it to $S^*$ since $a_4 = 1$, updating the explanation to $\{v_1, v_4\}$.  
Finally, at \textbf{step 5}, the agent considers $v_5$ and includes it in the explanation since $a_5 = 1$, resulting in the final CF$\boldsymbol{^2}$ set $\mathcal{S}^* = \{v_1, v_4, v_5\}$.  
The action selection at each step is determined by a policy network, which integrates the embeddings of the current CF$\boldsymbol{^2}$. This example demonstrates how our framework effectively reduces the original session to its key components through a sequential decision-making process.

\noindent\textbf{Learning and Optimization.}
Given the MDP environment described, we aim to learn a stochastic policy \(\pi\) that maximizes the expected cumulative reward. This is achieved using the REINFORCE algorithm, which optimizes the policy parameters \(\theta\) to maximize the total expected reward, defined as:
\begin{equation}\footnotesize
\mathcal{L}_r = -Q(\theta),
\end{equation}
where \(Q(\theta)\) is the expected cumulative reward:
\begin{equation}\footnotesize
Q(\theta) = \mathbb{E}_{\pi} \left[ \sum\nolimits_{t=0}^{T-1} \gamma^t R(s_{t+1}, a_{t+1}) \right],
\end{equation}
where \( \gamma \) is the discount factor, \( R(s_{t+1}, a_{t+1}) \) is the reward at step \( t+1 \), and \( \theta \) denotes the set of policy parameters.
The gradient of the expected cumulative reward with respect to the policy parameters \(\theta\) is given by:
\begin{equation}\footnotesize
    \nabla_{\theta} J(\theta) = \mathbb{E}_{\pi} \left[ \sum\nolimits_{t=0}^{T-1} \nabla_{\theta} \log \pi(a_t \mid s_t; \theta) Q(s_t, a_t) \right].
\end{equation}
We iteratively optimize the policy parameters $\theta$ to maximize cumulative rewards, enabling the agent to generate high-quality factual and counterfactual explanations.

\subsection{Recommendation Enhancement Module}
We aim to enhance recommendation accuracy by fusing both factual explanations (FEs) and counterfactual explanations (CFEs). FEs serve as high-quality positive samples, while CFEs act as negative samples. By using these signals, we can fundamentally improve the SR model through contrastive learning, inspired by \cite{Xie2022ContrastiveLF}.

This approach ensures that the SR model $f_R$ is trained to achieve two main objectives:
(1) \textbf{Closeness of Positive Samples}: The model is trained to bring the representation for factual and counterfactual explanations $\mathbf{e}_{\mathcal{S}^*}$ ($\mathbf{e}_{\mathcal{S}^*},\mathbb{I}_{\mathcal{S}^*}^K=f_{R} (\mathcal{S}^*)$) closer to the representation of original session items $\mathbf{e}_{\mathcal{S}}$. This is because the recommendation generated from the explanation for the next item should be consistent with the recommendation derived from the original session (factual condition).
(2) \textbf{Separation of Negative Samples}: The model is trained to increase the distance between the representation of the original session and the session without the explanation items $\mathbf{e}_{\mathcal{S} \setminus \mathcal{S}^*}$($\mathbf{e}_{\mathcal{S} \setminus \mathcal{S}^*},\mathbb{I}_{\mathcal{S} \setminus \mathcal{S}^{*}}^K=f_{R} (\mathcal{S} \setminus \mathcal{S}^*)$). This is because removing these items based on the counterfactual condition would lead to different recommendation results (counterfactual condition). 

Contrastive learning involves modifying the recommendation model's loss function to integrate insights from these explanations. Accordingly, as discussed in \cite{oord2018representation}, in SR we apply a contrastive loss function to differentiate positive samples from negative samples and maximize their separation, which is defined as:
\begin{equation}
\scriptsize
\mathcal{L}_{c} = - \sum_{i=1}^N\log \frac{\exp \left(\operatorname{sim}\left(\mathbf{e}_{\mathcal{S}}^{i}, \mathbf{e}_{\mathcal{S}^*}^{i}\right) \right)}{\exp \left(\operatorname{sim}\left(\mathbf{e}_{\mathcal{S}}^{i}, \mathbf{e}_{S^*}^{i}\right) \right) + \sum\limits_{k=1}^N \exp \left(\operatorname{sim}\left(\mathbf{e}_{\mathcal{S}}^{i}, \mathbf{e}_{\mathcal{S} \setminus \mathcal{S}^*}^{k}\right) \right)}, \end{equation}
where $N$ represents the number of session samples. $\mathbf{e}_{\mathcal{S}}^{i}$ represents the embedding of original sample. $ \mathbf{e}_{\mathcal{S}^*}^{i}$ is the embedding of positive sample (using CF$\boldsymbol{^2}$) and $\mathbf{e}_{\mathcal{S} \setminus \mathcal{S}^*}^k $ is the embedding of negative sample. The similarity function $sim()$ represents the cosine similarity between two embeddings.
The overall loss function guiding the fine-tuning of our SR model consists of two parts: the cross-entropy recommendation loss function $\mathcal{L}_{\text{rec}}$ and the contrastive loss function $\mathcal{L}_{\text{c}}$, 
\begin{equation}\label{eq:overallloss}\small
\mathcal{L}_{\text{fine-tune}} = \mathcal{L}_{\text{rec}} + \lambda \mathcal{L}_{\text{c}},
\end{equation}
where $\lambda$ is a hyper-parameter to balance the two losses; and the cross-entropy recommendation loss function is given by:

\begin{equation}\label{eq:overallloss}\small
\mathcal{L}_{\text{rec}} = -\sum\nolimits_{j}^{|\mathcal{V}|}\left(y_{j} \log \left(\hat{y}_{j}\right)\right),
\end{equation}
where  $y_{j}$  is the ground-truth label of item $v_{j}$ (either 1  or 0)  and $\hat{y}_{j}$  is the predicted score generated by the SR model $f_R$.

\subsection{Discussion on Optimality and Feasibility}
In our implementation, we carefully design the reward function to enforce the factual and counterfactual conditions of the optimization problem. This reward structure guides the RL agent toward selecting item subsets that satisfy the prescribed explanatory criteria. Given the NP-hard nature of the problem and the exponential time complexity for exact solutions, we adopt RL as an approximate method. For a session with $\mathcal{|S|}$ items, our state space is $2^\mathcal{|S|}$, where at each time step, the agent makes a binary decision (select or not select) for each item, resulting in an action space of size $2$. The episode length equals the session length. Although the state space remains exponential, the RL approach avoids exhaustive state enumeration, effectively reducing the computational complexity from exponential to polynomial time. While this approach does not guarantee global optimality, it efficiently generates near-optimal solutions, achieving a balance between computational efficiency and explanation quality.

\section{Experiments}
We conduct extensive experiments to validate the effectiveness of our model and address five research questions (RQs)\footnote{Our code is at \url{https://anonymous.4open.science/r/FCESR_code-67EA}.}. \textbf{RQ1}: How does FCESR improve recommendation accuracy for SR? \textbf{RQ2}: How effective is FCESR at generating explanations? \textbf{RQ3}: How do various components of FCESR affect its overall performance? \textbf{RQ4}: How do different hyper-parameters impact the performance of FCESR? \textbf{RQ5}: How does the computational performance of FCESR vary across datasets?

\subsection{Experimental Settings}
\textbf{Datasets:} We adopt three datasets: Beauty, Baby, and Movielens-1Ms, each representing different consumer product categories from e-commerce platforms. For the Amazon datasets (Beauty and Baby\footnote{cseweb.ucsd.edu/~jmcauley/datasets.html.}), the last session is assigned to the test set, with the remaining sessions used for training    \cite{wu2023generic}. Additionally, 10\% of the training set is set aside as a validation set for model tuning. For MovieLens\footnote{grouplens.org/datasets/movielens/.}, 75\% of the data is used for training, 15\% for validation, and the remaining 10\% serves as the test set to ensure balanced data partitioning. We split user interactions into sessions by day and exclude items with more than 100 interactions and sessions shorter than two interactions. This preprocessing strategy allows us to validate FCESR's capability to generate explanations effectively across various session lengths. Table \ref{table1:dataset_statistics} shows the data statistics.

\begin{table}[t]
\caption{Dataset statistics.}
\vspace{-0.1in}
\centering
\renewcommand{\arraystretch}{0.9}
\begin{tabular}{|l|c|c|c|}
\toprule
dataset & Beauty & Baby & MovieLens \\
\midrule
\#training sessions & 14,736 & 12,691 & 58,410 \\
\#validation sessions & 1,638 & 1,411 & 11,683 \\
\#testing sessions & 1,499 & 1,396 & 7,788 \\
\#average training session length & 6.64 & 5.66 & 9.47 \\
\#items & 8104 & 5267 & 3953 \\
\bottomrule
\end{tabular}
\label{table1:dataset_statistics}
\vspace{-0.1in}
\end{table}

\begin{table*}[htbp]
\caption{Overall accuracy comparison on the three datasets.}
\vspace{-0.1in}
\centering
\fontsize{9}{11}\selectfont
\adjustbox{max width=\textwidth}{
\begin{tabular}{|c|c|c|c|c|c|c|c|c|c|c|c|c|c|c|c|c|c|c|c|}
\hline
\multirow{2}{*}{Dataset} & \multirow{2}{*}{Metric} 
& \multicolumn{3}{c|}{GRU4REC} 
& \multicolumn{3}{c|}{NARM} 
& \multicolumn{3}{c|}{SRGNN} 
& \multicolumn{3}{c|}{GCSAN}  
& \multicolumn{3}{c|}{BERT4REC}
& \multicolumn{3}{c|}{COCO-SBRS} \\
\cline{3-20}
& & Base & FCESR & Improv. 
  & Base & FCESR & Improv. 
  & Base & FCESR & Improv. 
  & Base & FCESR & Improv. 
  & Base & FCESR & Improv. 
  & Base & FCESR & Improv. \\
\hline
\multirow{5}{*}{Baby} 
& HR@5   & 1.50 & 2.01 & 33.33\%** & 2.15 & 2.94 & 36.67\%** & 1.36 & 1.65 & 21.05\%** & 1.36 & 1.77 & 29.82\%** & 2.01 & 2.36 & 17.86\%** & 3.10 & 3.43 & 10.65\%** \\
& HR@10  & 2.72 & 3.22 & 18.42\%** & 3.37 & 3.80 & 12.77\%** & 2.44 & 3.58 & 47.06\%** & 2.58 & 2.77 & 7.41\%**  & 3.44 & 3.65 & 6.25\%**  & 4.50 & 4.62 & 2.67\% \\
& HR@20  & 4.80 & 5.37 & 11.94\%** & 4.73 & 5.80 & 22.73\%** & 4.15 & 5.44 & 31.03\%** & 4.37 & 4.47 & 2.19\%*   & 4.58 & 5.16 & 12.49\%** & 6.20 & 6.25 & 0.81\% \\
& NDCG@5 & 0.92 & 1.11 & 21.56\%** & 1.61 & 1.96 & 22.02\%** & 0.80 & 0.99 & 23.93\%** & 0.84 & 1.10 & 30.98\%** & 1.23 & 1.39 & 13.20\%** & 2.10 & 2.32 & 10.48\%** \\
& NDCG@10& 1.32 & 1.51 & 14.27\%** & 1.99 & 2.25 & 12.66\%** & 1.14 & 1.62 & 41.50\%** & 1.24 & 1.41 & 14.22\%** & 1.69 & 1.71 & 0.89\%    & 2.60 & 2.75 & 5.77\%** \\
& NDCG@20& 1.84 & 2.04 & 10.79\%** & 2.33 & 2.75 & 17.66\%** & 1.57 & 2.08 & 32.49\%** & 1.68 & 1.83 & 9.09\%**  & 1.97 & 2.15 & 8.95\%**  & 3.00 & 3.13 & 4.33\%** \\

\hline

\multirow{6}{*}{Beauty} 
& HR@5   & 19.15 & 19.18 & 0.17\%   & 20.01 & 21.04 & 5.11\%** & 9.01  & 10.14 & 12.59\%** & 10.20 & 10.24 & 0.33\%   & 12.07 & 14.74 & 22.10\%** & 9.49 & 9.79 & 3.21\%** \\
& HR@10  & 22.01 & 22.21 & 0.91\%   & 22.62 & 23.50 & 3.93\%** & 11.47 & 12.81 & 11.67\%** & 14.41 & 14.48 & 0.46\%   & 14.41 & 17.28 & 19.91\%** & 16.13 & 16.43 & 1.83\%** \\
& HR@20  & 25.15 & 25.22 & 0.27\%   & 24.48 & 25.73 & 5.09\%** & 14.01 & 15.61 & 11.43\%** & 18.21 & 18.51 & 1.65\%*  & 17.41 & 20.95 & 20.31\%** & 26.33 & 26.54 & 0.80\%* \\
& NDCG@5 & 15.00 & 15.09 & 0.62\%   & 16.93 & 17.11 & 1.10\%*  & 7.03  & 7.89  & 12.23\%** & 8.67  & 8.96  & 4.32\%** & 8.97  & 11.63 & 29.61\%** & 5.64 & 5.78 & 2.51\%** \\
& NDCG@10& 15.96 & 16.10 & 0.88\%   & 17.77 & 17.91 & 0.76\%   & 7.83  & 8.75  & 11.72\%** & 8.24  & 8.52  & 3.43\%** & 9.72  & 12.46 & 28.15\%** & 7.78 & 7.81 & 0.45\%* \\
& NDCG@20& 16.75 & 16.85 & 0.60\%   & 18.24 & 18.47 & 1.25\%*  & 8.47  & 9.47  & 11.77\%** & 9.20  & 9.57  & 3.95\%** & 10.47 & 13.39 & 27.85\%** & 10.35 & 10.46 & 1.06\%** \\

\hline

\multirow{6}{*}{MovieLens} 
& HR@5   & 7.74 & 8.68 & 12.14\%** & 10.46 & 11.69 & 11.77\%** & 6.75 & 7.46 & 10.44\% & 4.73 & 4.91 & 3.87\%** & 6.39 & 7.10 & 11.00\%** & 8.20 & 8.67 & 5.70\%** \\
& HR@10  & 11.65 & 13.06 & 12.10\%** & 15.88 & 16.79 & 5.74\%**  & 11.28 & 12.46 & 10.49\%** & 7.35 & 7.51 & 2.14\%* & 9.60 & 10.18 & 5.97\%** & 14.63 & 14.93 & 2.02\%** \\
& HR@20  & 16.63 & 18.13 & 9.02\%**  & 23.04 & 23.50 & 2.01\%*   & 17.64 & 19.77 & 12.09\%** & 11.41 & 11.50 & 0.82\%  & 14.58 & 15.08 & 3.46\%** & 24.46 & 24.78 & 1.33\%** \\
& NDCG@5 & 5.11  & 6.46  & 26.42\%** & 7.19  & 7.32  & 1.85\%*   & 4.28  & 4.90  & 14.53\%** & 3.13 & 3.29 & 5.24\%** & 4.19 & 4.59 & 9.44\%** & 5.00 & 5.21 & 4.22\%** \\
& NDCG@10& 6.37  & 7.14  & 12.09\%** & 8.93  & 9.55  & 6.94\%**  & 5.74  & 6.51  & 13.41\%** & 3.97 & 4.13 & 3.98\%** & 5.23 & 5.43 & 3.87\%** & 7.06 & 7.23 & 2.43\%** \\
& NDCG@20& 7.63  & 8.83  & 15.73\%** & 10.73 & 11.92 & 11.08\%** & 7.33  & 8.34  & 13.81\%** & 4.99 & 5.60 & 5.13\%** & 6.48 & 6.59 & 1.80\%*  & 9.52 & 9.87 & 3.64\%** \\

\hline
\end{tabular}
}
\label{tab:performance_comparison}
\end{table*}

\noindent\textbf{Baseline Models:} 
Our work aims to provide explanations for existing SR models while leveraging the generated explanations as positive and negative samples for data augmentation to enhance their accuracy. Therefore, our baselines consist of two parts: \textbf{explanation models} for validating FCESR's interpretability, and \textbf{SR models} for evaluating the effectiveness of our explanation-based data augmentation approach.
For SR models, we consider several state-of-the-art backbones: \textbf{GRU4REC} \cite{hidasi2015session} uses GRU to capture sequential interactions. \textbf{NARM} \cite{li2017neural} uses an attention mechanism to better capture evolving session interests. \textbf{BERT4REC} \cite{sun2019bert4rec} employs bidirectional self-attention with a Cloze task for masked item prediction. \textbf{SR-GNN} \cite{wu2019session} utilizes a gated GCN with self-attention for session representation. \textbf{GCSAN} \cite{xu2019graph} combines GNNs with self-attention to address both short- and long-term session dependencies. \textbf{COCO-SBRS} \cite{wang2023coco} models outer-session confounders (OSCs) to capture causal user intent.


Regarding explanation models, as the first work to propose both factual and counterfactual explanations in SR, we consider two baseline approaches for explanation generation: (1) \textbf{Random Selection}, a simple baseline that randomly samples items from the session as explanations. (2) \textbf{CRVAE} \cite{xu2021learning}, a sequential recommendation explanation model that generates true explanations utilizing a VAE structure. Since no official implementation is available, we implemented it based on the pseudo-code provided in the original paper. We also note that \cite{cfe2024} does not explicitly generate counterfactual explanations but only outputs importance scores and rankings for items, and \cite{li2024attention} does not provide reproducible source code. We attempted to reproduce their methods but were unable to obtain comparable implementations under the SR setting. Therefore, both methods are excluded from comparison.


\noindent\textbf{Evaluation Metrics:} We use  Hit Ratio (HR@K) and Normalized Discounted Cumulative Gain (NDCG@K) as evaluation metrics. HR@K measures the percentage of test instances in which the target item appears in the top-$K$ list, while NDCG@K also considers the ranking quality, giving higher weight to correct items that are positioned higher in list. We set $\{K=5, 10, 20\}$ for a comprehensive evaluation, with higher values indicating better performance.

Regarding explanations, a good explanation should meet both factual and counterfactual conditions, meaning that the items included in the explanation should significantly influence the recommended results.
We introduce the concepts of Probability of Sufficiency (PS) and Probability of Necessity (PN), derived from causal inference theory \cite{pearl2016causal} and used in GNN explanation task \cite{tan2022learning}. They quantitatively assess the quality of explanations. In SR, we use them to evaluate whether the items generated by our method contribute to the final recommendation outcomes. Let $\mathbb{I}_{\mathcal{S}}^K$ be the set of items in the recommendation list. We define the PS and PN as:
\begin{equation}\footnotesize
    \mathrm{PS}=\frac{\sum \mathbf{1}_{\mathbb{I}_{\mathcal{S}}^K}(i_{\mathcal{S}^*}^*)}{|\mathcal{S}|}, \quad \mathrm{PN}=\frac{\sum \mathbf{1}_{\mathbb{I}_{\mathcal{S}}^K}^c(i_{\mathcal{S}^*}^*)}{|\mathcal{S}|},
\end{equation}
where $i_{\mathcal{S}^*}^*$ is the next item prediction for the found explanation $\mathcal{S}^*$, $\mathbf{1}_{\mathbb{I}_{\mathcal{S}}^K}(i_{\mathcal{S}^*}^*)$ and $\mathbf{1}_{\mathbb{I}_{\mathcal{S}}^K}^c(i_{\mathcal{S}^*}^*)$ are indicator functions for $i_{\mathcal{S}^*}^* \in \mathbb{I}_{\mathcal{S}}^K$ and $i_{\mathcal{S}^*}^* \notin \mathbb{I}_{\mathcal{S}}^K$, respectively. These metrics assess the sufficiency and necessity of explanation items in maintaining or altering the recommendation model's predictions. Higher values indicate more effective explanations. 
To combine these metrics, we calculate the $F_{ns}$ score as $F_{ns}=\frac{2 \cdot P N \cdot P S}{P N+P S}$. This score emphasizes a balance between sufficiency and necessity in our explanations.

\subsection{Recommendation Performance (RQ1)}
 The comparative results on five non-explainable SR methods
 are shown in Table \ref{tab:performance_comparison}. The following observations are noted.
 (1) FCESR substantially improves recommendation accuracy across all datasets and most evaluation metrics, showing remarkable enhancements compared to baseline models.
 (2) BERT4Rec and GRU4Rec exhibit significant performance improvements across all scenarios, whereas GCSAN's improvements are relatively modest. This can be partially explained by the data in Table \ref{tab:explanation_results}, which shows that BERT4Rec and GRU4Rec excel in generating effective explanations across all datasets. BERT4Rec, for example, achieves up to 79.67\% effectiveness on MovieLens. This ability to produce high-quality explanations results in valuable samples that boost accuracy after fine-tuning. In contrast, GCSAN's lower explanation effectiveness likely contributes to its smaller accuracy gains. These findings underscore that high-quality explanations not only enhance interpretability but also provide crucial training samples, thereby improving overall recommendation accuracy. This highlights the connection between explanation quality and recommendation effectiveness.


\begin{table}[htbp]
\centering
\vspace{-1em}
\caption{Comparative results regarding top-$1$ explanation on the three datasets. The best is boldfaced.}
\label{tab:explanation_results}

\setlength{\tabcolsep}{1.5pt} 
\renewcommand{\arraystretch}{1.1}

\resizebox{\columnwidth}{!}{
\begin{tabular}{|c|c|cc>{\columncolor{gray!20}}c|cc>{\columncolor{gray!20}}c|cc>{\columncolor{gray!20}}c|}
\hline
\multirow{2}{*}{\textbf{Data}} & \multirow{2}{*}{\textbf{Models}} & \multicolumn{3}{c|}{\textbf{Random}} & \multicolumn{3}{c|}{\textbf{CRVAE}} & \multicolumn{3}{c|}{\textbf{FCESR}} \\
\cline{3-11}
 & & PN & PS & $F_{ns}$ & PN & PS & $F_{ns}$ & PN & PS & $F_{ns}$ \\
\hline
\multirow{6}{*}{\textbf{Baby}} 
 & NARM     & 99.95\% & 0.07\% & 0.14\% & 98.83\% & 23.10\% & 37.45\% & 99.91\% & 50.66\% & \textbf{67.23\%} \\
 & SRGNN    & 99.78\% & 0.21\% & 0.42\% & 99.46\% & 1.58\% & 3.11\% & 99.77\% & 2.12\% & \textbf{4.15\%} \\
 & GRU4REC  & 99.91\% & 0.04\% & 0.08\% & 98.83\% & 54.67\% & 70.40\% & 99.76\% & 96.51\% & \textbf{98.11\%} \\
& GCSAN    & 99.59\% & 0.28\% & 0.56\% & 99.47\% & 0.60\% & 1.19\% & 99.86\% & 1.20\% & \textbf{2.37\%} \\
 & BERT4REC & 99.92\% & 0.12\% & 0.24\% & 98.14\% & 19.01\% & 31.85\% & 98.26\% & 53.13\% & \textbf{68.97\%} \\
 & COCOSBRS & 99.61\% & 9.40\% & 17.18\% & 98.90\% & 18.42\% & 31.06\% & 98.65\% & 32.88\% & \textbf{49.32\%} \\
\hline
\multirow{6}{*}{\textbf{Beauty}} 
 & NARM     & 99.94\% & 0.07\% & 0.14\% & 99.35\% & 29.85\% & 45.91\% & 99.27\% & 73.48\% & \textbf{84.45\%} \\
 & SRGNN    & 99.95\% & 0.03\% & 0.06\% & 98.32\% & 15.92\% & 27.40\% & 99.35\% & 18.39\% & \textbf{31.04\%} \\
 & GRU4REC  & 99.89\% & 0.04\% & 0.08\% & 99.24\% & 56.62\% & 72.10\% & 99.63\% & 98.51\% & \textbf{99.07\%} \\
& GCSAN    & 99.95\% & 0.05\% & 0.10\% & 98.30\% & 6.17\% & 11.61\% & 99.11\% & 7.34\% & \textbf{13.67\%} \\
 & BERT4REC & 99.90\% & 0.09\% & 0.18\% & 99.78\% & 18.20\% & 30.78\% & 98.76\% & 19.92\% & \textbf{33.15\%} \\
 & COCOSBRS & 99.40\% & 0.07\% & 0.14\% & 99.93\% & 23.86\% & 38.52\% & 99.42\% & 31.27\% & \textbf{47.57\%} \\
\hline
\multirow{6}{*}{\textbf{MovieLens}} 
 & NARM     & 98.92\% & 0.11\% & 0.22\% & 98.05\% & 24.77\% & 39.55\% & 99.58\% & 33.45\% & \textbf{50.08\%} \\
 & SRGNN    & 99.77\% & 0.27\% & 0.54\% & 99.31\% & 19.01\% & 31.91\% & 99.49\% & 28.63\% & \textbf{44.46\%} \\
 & GRU4REC  & 99.15\% & 0.57\% & 1.13\% & 99.11\% & 41.89\% & 58.89\% & 99.56\% & 92.46\% & \textbf{95.88\%} \\
 & GCSAN    & 99.42\% & 0.09\% & 0.18\% & 98.15\% & 38.46\% & \textbf{55.26\%} & 98.43\% & 32.49\% & 48.85\% \\
 & BERT4REC & 99.63\% & 0.07\% & 0.14\% & 98.90\% & 36.49\% & 53.30\% & 99.74\% & 79.67\% & \textbf{88.58\%} \\
 & COCOSBRS & 98.37\% & 0.05\% & 0.10\% & 99.50\% & 33.10\% & 49.68\% & 99.46\% & 61.32\% & \textbf{75.87\%} \\
\hline
\end{tabular}
}
\vspace{-1em}
\end{table}

\subsection{Explanation Performance (RQ2)}
In FCESR, both factual and counterfactual reasoning are used to generate explanations. The results are presented in Table \ref{tab:explanation_results}. We compare our results with randomly generated explanations and CRVAE for each method on different datasets and observe that FCESR successfully generates explanations for most of the recommended items, with following key findings:
 (1) FCESR consistently outperforms both Random and CRVAE baselines across most scenarios. (2) High PN values are observed across all models and datasets, likely due to the relative ease of generating counterfactual data by simply altering the predicted next item. (3) SRGNN and GCSAN exhibit lower PS values, particularly on the Baby dataset (PS: 2.12\% and 1.2\% respectively). This may be attributed to the reliance of SRGNN and GCSAN on consecutive transitions for graph construction, as removing items destroys this critical adjacency structure. (4) BERT4REC and GRU4REC show higher PS values, likely because of their superiority to capture long-term dependencies and contextual information in time series. BERT4REC's bidirectional Transformers and GRU4REC's RNN architecture excel at modeling complex sequential data, leading to more accurate and abundant explanations in challenging recommendation tasks.

\begin{figure}[htbp]
\centering
\includegraphics[width=0.45\textwidth]{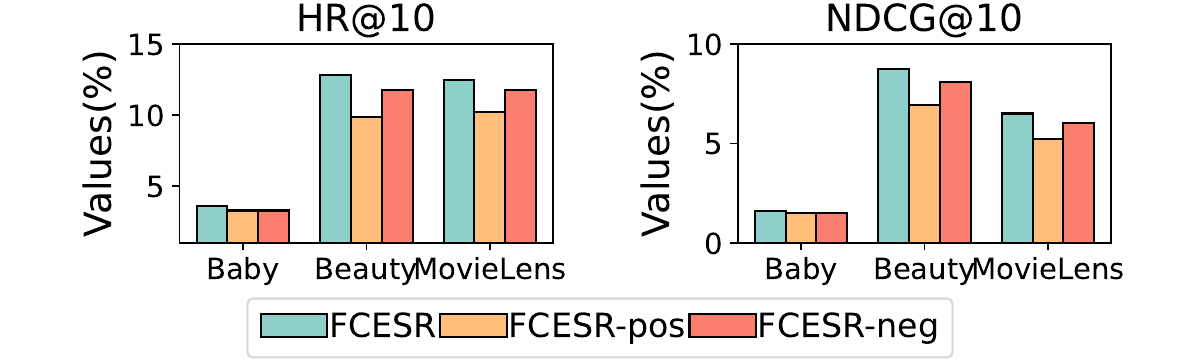}
\vspace{-0.1in}
\caption{Impact of Positive and Negative Samples.}
\label{fig:cons ablation}
\vspace{-0.15in}
\end{figure}

\begin{table}[htbp]
\caption{Impact of different reward functions.}
\vspace{-0.1in}
\centering
\addtolength{\tabcolsep}{-1pt}
\footnotesize
\begin{tabular}{|l|c|cccc|}
\toprule Dataset & Variant & PN & PS & $F_{ns}$ & avg\_len \\ 

\midrule
\multirow{3}{*}{Beauty} 
& FCESR (w/o) spars & 99.78\% & 98.79\% & 99.28\% & 6.62\\
& FCESR (w/o) fe & 98.25\% & 88.21\% & 92.96\% & 5.66 \\
& FCESR & 99.63\% & 98.51\% & 99.07\% & 5.89 \\
\midrule
\multirow{3}{*}{Baby} & FCESR (w/o) spars & 98.68\% & 98.79\% & 98.73\% & 5.64 \\ 
& FCESR (w/o) fe & 98.25\% & 88.23\% & 92.97\% & 5.63 \\ 
& FCESR & 99.76\% & 96.51\% & 98.11\% & 5.58 \\

\bottomrule
\end{tabular}
\label{table4:recommendation_performance}
\vspace{-0.15in}
\end{table}

\subsection{Ablation Study (RQ3)}
We evaluate the key designs of FCESR, namely complex reward functions and the use of positive and negative samples. For the reward function, we examine three variants: FCESR without sparsity reward and FCESR with only factual reward. Experiments are conducted on the Baby and Beauty datasets, with results using GRU4REC reported in Table \ref{table4:recommendation_performance}. Similar trends are observed across other backbone models. The findings show that each component contributes significantly to the overall performance, demonstrating the robustness of the reward design.

The results illustrate a trade-off between completeness and conciseness. On Baby, removing the feature entropy reward (w/o fe) notably degrades performance ($F_{ns}$ drops to 92.97\%), while omitting the sparsity reward (w/o spars) results in a longer average length (5.64). The complete FCESR achieves the optimal balance, maintaining high accuracy ($F_{ns}$ 98.11\%) with the shortest average length (5.58). A similar pattern is observed on Beauty, where the variant without sparsity reward produces the longest explanations, whereas the complete model significantly reduces length without compromising effectiveness. These results confirm the necessity of combining multiple reward components.

Although simplified variants may excel on individual metrics, the complete FCESR offers the best trade-off between explanation quality, conciseness and comprehensiveness.

\begin{figure}[htbp]
\centering
\includegraphics[width=0.45\textwidth]{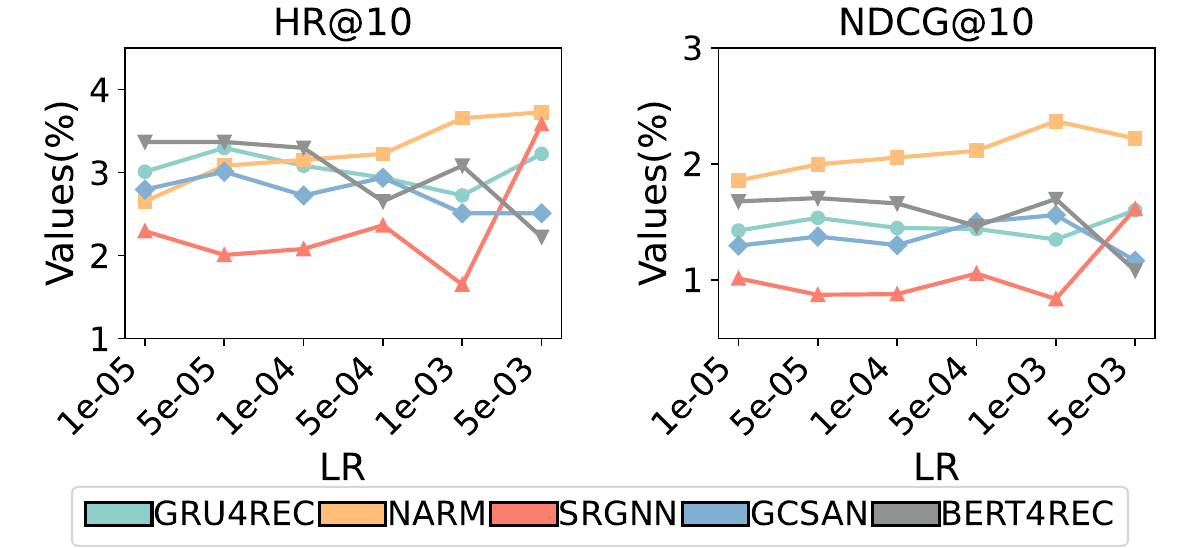}
\vspace{-0.15in}
\caption{Impact of different lr.}
\label{fig:lr}
\vspace{-0.1in}
\end{figure}

To evaluate the impact of positive and negative samples, we consider two variants: \textbf{FCESR-pos}, which only uses positive samples in the recommendation enhancement module, and \textbf{FCESR-neg}, which only uses negative samples. Figure \ref{fig:cons ablation} presents the experimental results on three datasets using the SRGNN model. Similar trends are observed across other SR models. The results validate that each component contributes to the final performance, with the highest improvement in recommendation accuracy achieved when both positive and negative samples are utilized.

\begin{figure}[htbp]
\centering
\includegraphics[width=0.45\textwidth]{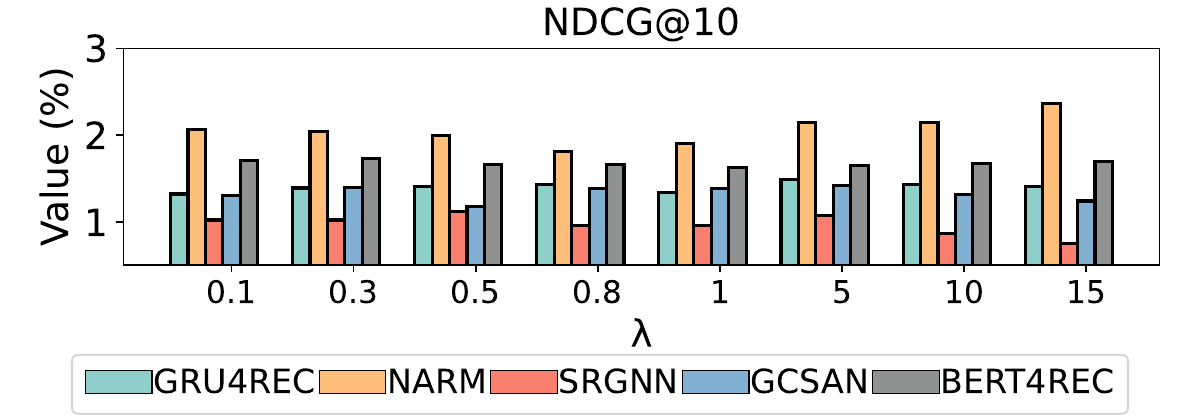}
\vspace{-0.15in}
\caption{Impact of different $\lambda$.}
\label{fig:lambda}
\vspace{-0.15in}
\end{figure}

\subsection{Sensitivity of Hyper-parameters (RQ4)}
We further conduct analysis of two key hyperparameters during the fine-tuning phase, i.e., learning rate (lr) and contrastive loss coefficient ($\lambda$), explored a range of values [0.00001, 0.0005, 0.001, 0.005, 0.01], and [0.1, 0.3, 0.5, 0.8, 1, 5, 10, 15], respectively.
The results on Baby are shown in Figures \ref{fig:lr} and \ref{fig:lambda}. As demonstrated, FCESR demonstrates considerable robustness to hyperparameter variations, as evidenced by the relatively modest changes in accuracy across different settings. This insensitivity to hyperparameter adjustments further validates the stability of our FCESR.

\subsection{Time Complexity Analysis (RQ5)}
The experiments were conducted on a computational infrastructure running Ubuntu 18.04, with 8 CPU cores and 64GB of memory. 
Across the three datasets, the average runtime per episode is approximately 115.6 seconds on Baby, 218.2 seconds on Beauty, and 253.2 seconds on MovieLens. The total training time varies depending on the stopping criteria (reward convergence, network convergence, or maximum episode limit). The runtime performance exhibits correlations with both the average session length and the total number of sessions in each dataset, as our framework needs to traverse all items within a session during each episode. For instance, the Baby dataset exhibits notably shorter runtime compared to the other two datasets, primarily attributed to its shorter session length. Overall, the computational complexity remains moderate, with session length being the dominant factor affecting runtime performance. 

\section{Conclusions}
We introduce the FCESR framework, a model-agnostic plug-in that addresses the explainability issue in existing SR while also improving recommendation accuracy. Our key innovation lies in formulating the generation of factual and counterfactual explanations as a combinatorial optimization problem. We tackle this challenge through a novel framework that uses reinforcement learning for capturing sequential dependencies and balancing factual and counterfactual explanations. We further leverage contrastive
learning to utilize the generated explanations for improving
recommendation accuracy. Extensive experiments comparing against two explanation baselines on five advanced SR models across three datasets demonstrate FCESR's superior performance in both improving recommendation accuracy and generating high-quality explanations. Future work could focus on adapting the FCESR framework to different recommendation scenarios and conducting large-scale user studies to assess the impact of these explanations on real users' decision-making and satisfaction.




\bibliographystyle{ACM-Reference-Format}
\bibliography{ref}

\end{document}